\documentclass[aps,prl,reprint,groupedaddress]{revtex4-1}
\usepackage{amsmath,amssymb}
\usepackage{graphicx}
\usepackage{units}
\usepackage{latexsym}
\usepackage{bm}
\usepackage{color}
\usepackage{epstopdf}
\usepackage{enumitem}
\newcommand{\subscript}[2]{$#1 _ #2$}

\begin{document}

	\title{
		\large
		Experimental Evidence for Selection Rules in \\
		Multiphoton Double Ionization of Helium}
	
	\author{
		K. Henrichs$^1$, S. Eckart$^1$, A. Hartung$^1$, D. Trabert$^1$, J. Rist$^1$, H. Sann$^1$, M. Pitzer$^2$,\\ M. Richter$^1$, H. Kang$^{1,3}$, M. S. Sch\"offler$^1$, M. Kunitski$^1$, T. Jahnke$^1$} \author{R. D\"orner$^1$}\email{doerner@atom.uni-frankfurt.de}

	\affiliation{$^1$Institut f\"ur Kernphysik, J.~W.~Goethe-Universit\"at, Max-von-Laue-Str. 1, 60438 Frankfurt am Main, Germany\\
		$^2$Institut f\"ur Physik, Universit\"at Kassel, Heinr.-Plett-Str. 40, 34132 Kassel, Germany\\
		$^3$State Key Laboratory of Magnetic Resonance and Atomic and Molecular Physics, Wuhan Institute of Physics and Mathematics, Chinese Academy of Sciences, Wuhan 430071, China}
	
	%\begin{abstract}
	
	%\end{abstract}
	\date{\today}
	\begin{abstract}
		We report on the observation of a multi peak structure in the correlated two electron energy distribution from strong field double ionization of helium using laser pulses with a wavelength of \unit[394]{nm} and an intensity of $\unit[3\cdot10^{14}]{W/cm^2}$. For selected regions of electron emission angles the peaks emerging at energies corresponding to an odd number of absorbed photons are suppressed. We interpret this as direct results of quantum mechanical selection rules. For Neon these features occur for even photon number. By that we attribute this effect to the parity of the continuum wave function. A comparison with analogous data for single photon double ionization is given.
	\end{abstract}
	
	\maketitle
	
	The ionization of atoms by an electromagnetic wave has been studied across a tremendous range of light intensities and wavelengths. Our physical picture of the interaction process changes completely from the perturbative single photon regime considered by Einstein over the multiphoton to the tunneling regime accessible today with table top lasers and free electron laser sources. Despite the very different interaction mechanisms, there are still several quantum features, which are expected to be universal across these different intensity and wavelength regimes. The most obvious one is the quantization of energy transfer from the field to the ejected electron(s). Along with this energy quantization more subtle features are predicted to be quantized, such as the change of parity and angular momentum between the bound initial and the final state of the fragments (electrons and ion). The transfer of one quantum of energy, i.e. the absorption of one photon, is accompanied by a parity change between initial and final state. It is also, within the dipole approximation, accompanied by the transfer of one quantum of angular momentum.
	
	Energy quantization is easily observed in the case of {\em single ionization} in the different intensity regimes: For one photon processes it is the basis for spectroscopy. For multiphoton processes in the intensity regime of  \unit[$10^{13}-10^{14}$]{W/cm$^2$} it produces a comb of discrete structures in the energy spectrum of the emitted electron spaced by the photon's energy. These structures are commonly described as ATI (Above-Threshold-Ionization) peaks, referring to the fact that more photons can be absorbed than needed to overcome the ionization threshold. For {\em double ionization} the quantized excess energy can be shared between the electrons, but the energy quantization can still be observed in the sum energy of both electrons \citep{Wehlitz91} and is used accordingly for spectroscopy in the one photon case, as well \citep{Penent13,Penent03}. In double ionization in the multiphoton strong field regime the discretisation of the electron's sum energy ($E_1+E_2$) leads to ATDI (Above-Threshold-Double-Ionization) peaks at energies given by:
	\begin{eqnarray}
	E_1 + E_2 = n h\nu - I_{p1} - I_{p2} - 2U_p~.
	\label{eqndouble}
	\end{eqnarray}
	where $I_{p1,2}$ refers to the first and the second ionization potential of the atom and $U_{p}$ to the ponderomotive energy. The latter is defined in atomic units as $U_p=I/2c\omega_{Laser}^2$ ($c$ being the speed of light, $\omega_{Laser}$ the laser frequency and $I$ the laser intensity). $n$ refers to the number of absorbed photons. While eq.~\ref{eqndouble} has been confirmed by many calculations \citep{Lein01,Liao10,Parker06,Zielinski16,Parker01,Taylor11,Thumm14} most strong field double ionization experiments failed to resolve such discrete structures, presumably due the variation of $U_p$ across the laser focus (see \citep{HenrichsPRL13} for the only experimental confirmation).
	
\begin {figure}[t]
	\begin{center}
		\includegraphics[width=0.9\linewidth]{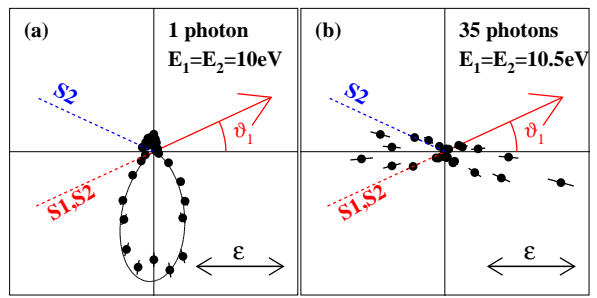}
		\caption{Joint angular distributions and selection rules of double ionization of helium (double ionization potential \unit[79]{eV}). (a) absorption of one linearly polarized photon ($h\nu =$ \unit[99]{eV}, synchrotron radiation, reanalysis of data from experiment reported in \cite{braeuning98}). Both electrons share the excess energy equally, electron 1 was selected at a \unit[20]{deg}$<\vartheta_1<$\unit[30]{deg} angle (red arrow) with respect to the polarization axis $\epsilon$ (horizontal). The angular distribution of the second electron measured in coincidence is plotted for electrons in the plane of the paper ($\pm$\unit[25]{deg}). The full line shows a parametrization of the cross section (eq. on page 5154 in Ref. \cite{braeuning98}). The angles blocked by selection rules S$_1$ and S$_2$ are shown by the red and blue dashed line. (b) Same geometry as (a) but for the case of absorption of 35 photons ($h\nu=$\unit[3.2]{eV}, \unit[$3\cdot10^{14}$]{W/cm$^2$}, \unit[45]{fs}). Only electrons of equal energy which occurred at energies depicted by the circle in Fig.\ref{fig2} are plotted.}
		\label{fig1}
	\end{center}
	\end {figure}

	The consequences of parity and angular momentum transfer to the two electron continuum has been first worked out in detail for the single photon case \citep{MaulBr95}. They lead to selection rules, which are observable as nodes in the differential cross section occurring for certain emission angles and energy sharings of both electrons if they are measured in coincidence (see \citep{Malcherek97} for a generalization). In the single photon regime these theoretical predictions have been confirmed also experimentally in detail, e.~g. for the double ionization of helium \citep{Briggs00jpb}, neon \citep{Kraessig96} and H$_2$ \citep{Gisselbrecht06,Weber04}. These selection rules have been generalized for absorption of more than one photon by Ni et al. \citep{Ni12}. They have formulated two rules, which hold for double ionization of an even initial state and an odd number of absorbed photons. For the special case of double ionization of helium the phase space covered by the first selection rule is a subset of the phase space covered by selection rule S$_2$:
	
	\begin{enumerate}[label=(\subscript{S}{{\arabic*}})]
		\item Back-to-back emission of electrons with equal energy ($k_1=k_2$): $\vec{k}_1=-\vec{k}_2$ is forbidden (red dashed line in Fig. \ref{fig1}).
		\item The emission of electrons of equal energy with angles $\vartheta_1 + \vartheta_2=180^o$ is forbidden (blue and red dashed lines in Fig. \ref{fig1}).
	\end{enumerate}
	
	Both rules result from the Pauli principle, the parity and total angular momentum of the two-body state. Accordingly, these selection rules do not depend on the physical mechanism of emission: they hold true for shake-off as well as knock-off \citep{Knapp02}; for sequential as well as nonsequential ionization; and for the regime of tunnelionization as well as for three (five and so on) photon transition at a free electron laser.
	In all quantum mechanical calculations, these selection rules are implicitly implemented (see e.g. in direct solutions of the time dependent Schroedinger equation \citep{Lein01,Zielinski16,Thumm15}, S-Matrix calculations \citep{Becker94,Faisal05} or close coupling methods \citep{Pindzola17}). Despite their universal character, however, these quantum mechanical selection rules have never been observed in experiments beyond the single photon case. The only experimental hint was reported in experimental work on two photon double ionization where the selection rules do not apply. There doubly charged helium ions with zero momentum were observed, which are suppressed by the selection rules in the one photon case \citep{Doerner96,Kurka10}.
	
	 In Fig.\ref{fig1} we illustrate the joint electron angular distributions for the cases of double ionization by one photon of \unit[99]{eV} (reanalysis of dataset presented in \cite{braeuning98}) and the corresponding case of absorption of 35 photons of \unit[3.2]{eV} at \unit[$3\cdot10^{14}$]{W/cm$^2$} from our current experiment. The physical mechanisms leading to double ionization in both cases are completely different. Accordingly, this difference manifests themselves in vastly differing angular distributions. After absorption of a single photon the electrons evolve freely and the angular distribution obtained at \unit[20]{eV} excess energy is shaped by the joint action of electron repulsion and the influence of the selection rules which effectively blocks an extended range of angles (see \cite{Briggs00jpb} for details). As only one unit of angular momentum is absorbed from the field also the individual electrons have comparably low angular momenta and the angular distribution shows no narrow peaks. In the strong field multiphoton case, on the contrary, much higher angular momenta are present allowing for narrow features in the angular distributions to form. The sharply directed electron emission along the polarization axis of the laser light is caused by the strong electric field of the laser, which drives the electrons along this direction. The nodes expected due to the selection rules are supposed to appear at the angles indicated by the dashed lines. Obviously, these are practically impossible to observe on these extremely narrow angular distributions. 
	 We therefore pursued the following route in order to achieve the goal of providing the missing experimental evidence for the selection rules in strong field physics: Instead of selecting a subset of data corresponding to a fixed even or odd number of absorbed photons from the total dataset and inspecting the angular distributions (as done in Fig. \ref{fig1}), we rather select ranges of emission angles for both electrons which are either suppressed or permitted by the selection rules and then inspect the electron sum energy spectrum (\ref{eqndouble}). By inspecting this sum energy histogram we can investigate whether the absorption of even/odd number of photons  is present or supressed.
	
	In our experiment, we used linearly polarized light at a wavelength of \unit[394]{nm} ($\hbar\nu\approx$\unit[3.2]{eV}) and employed a COLTRIMS Reaction Microscope \citep{Ullrich03} for the coincidence detection of all charged particles. A Ti:Sa laser system (Wyvern-500, KMLabs, \unit[45]{fs}, \unit[100]{kHz}) at a central wavelength of \unit[788]{nm} was used to generate the second harmonic at \unit[394]{nm} with a $\unit[200]{\mu m}$  BBO crystal. We determined the intensity of the resulting \unit[394]{nm} pulses to be $3\cdot10^{14}$~W/cm$^2$ in situ by the ponderomotive shift of the sum energy of electron and proton after dissociative ionization of H$_2$. The calibration of $U_p$ ($\approx$ \unit[4.5]{eV}) has been verified by examining the energies of the ATI peaks of helium single ionization during the whole experiment (approximately 140 hours of data acquisition). From this we estimated the accuracy of our intensity calibration to be better than $\pm 20~\%$.  The laser pulses were focused onto a supersonic gas jet target by a spherical mirror ($f$=\unit[60]{mm}). Electrons and ions were guided by a parallel electric (\unit[1.6]{V/cm}) and magnetic field (\unit[6.3]{Gauss}) towards two position- and time-sensitive multi channel plate detectors with three-layer delayline anodes for position read-out \citep{Jagutzki02}. The measurement was conducted at a count rate of \unit[5]{kHz} ions and \unit[15]{kHz} electrons. We used a $5~\mu m$ nozzle and piezo controlled collimators in order to adjust the width of the gas jet to be much smaller than the laser's Rayleigh length. We found a ratio of He$^{2+}$/He$^{1+}=3\cdot10^{-4}$ at $3\pm 0.6\cdot10^{14}$~W/cm$^2$. The background pressure in the interaction chamber was below $2\cdot10^{-11}$~mbar, which was essential to reduce the amount of H$_2^+$ overlapping in time-of-flight with the He$^{2+}$ ions.\\
	
	The dataset recorded contains approx. 70000 events where two electrons were measured in coincidence with the He$^{2+}$ ion and about a factor 6 more in which only one of the two electron was detected. For these events we obtain the momentum of the missing electron from the measured momenta of the ion and the other electron using momentum conservation. In this case about $30~\%$ false H$_2^+$ coincidences had to be subtracted.
	
	\begin {figure}[t]
	\begin{center}
		\includegraphics[width=0.9\linewidth]{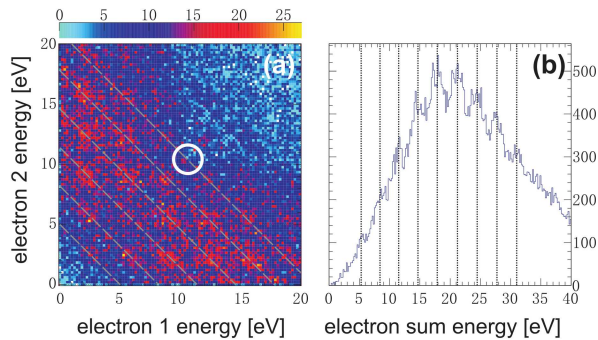}
		\caption{Energy spectra of \textit{two directly measured} electrons after double ionization of helium atoms by laser pulses with a peak intensity of $3\cdot10^{14}$~W/cm$^2$ and a central wavelength of \unit[394]{nm}. (a) Joint energy distribution of both electrons (energy of one electron plotted versus energy of the second). The grey diagonals lines indicate constant sum energies at the values calculated using eq. \ref{eqndouble}.
			The white circle at $E_1=E_2=$\unit[10.5]{eV} shows the region of events selected for Fig. \ref{fig1}b. (b) Sum energy of both electrons (same data as in a)). Vertical lines correspond to the grey lines in a) indicating the calculated position of the ATDI peaks.}
		\label{fig2}
	\end{center}
	\end {figure}
	
	In Fig. \ref{fig2}(b) we  show the sum energy of electrons measured in coincidence with a doubly charged He ion. The vertical lines represent the expected peak positions for the independently determined intensity of $3\cdot10^{14}$~W/cm$^2$ ($U_p \approx 4.5$~eV) according to eq.~\ref{eqndouble}. We used the lowest intensity possible to still reach sufficient count rates, since for higher intensities the ATDI peaks were not resolvable due to a stronger focal averaging effect. The measured peak positions perfectly match the calculated positions. This is the key to assign a total number of absorbed photons $n$ to each peak in the sum energy spectrum. Additionally, Fig. \ref{fig2}(a) shows the joint energy distribution of both emitted electrons. Grey diagonal lines represent the sum energies depicted by the vertical lines in Fig. \ref{fig2}(b). Along these diagonals distinct peaks occur. Such peaks have already been observed in the correlated energy spectrum in double ionization of argon \citep{HenrichsPRL13}. They were attributed to a doubly excited intermediate state. For helium similar peaks along the diagonal ATDI lines have been predicted in \citep{Zielinski16} for $800$~nm, the physical mechanism producing these peaks for helium remained open in these calculations.
	
	\begin {figure}[htbp]
	\begin{center}
		\includegraphics[width=0.95\linewidth]{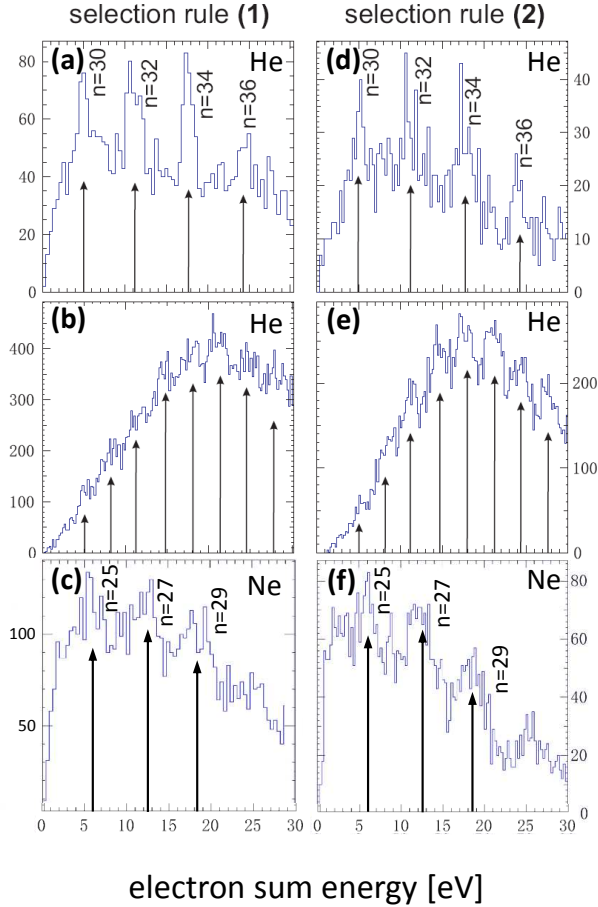}
		\caption{Demonstration of the consequences of quantum mechanical selection rules in double ionization of helium and neon. Left column: Selection rule S$_1$ (see text), which forbids $\vec{k}_1=\vec{k}_2$. The sum energy is plotted for all events with $140^o<\alpha<180^o$ ($\alpha$ is the angle between the two electrons) and (a), (c) equal energy sharing, (b) no equal energy sharing condition. In (a) and (c) equal energy sharing is fulfilled, if the energy of both electrons differs by less than $1~eV$. Right column: Selection rule S$_2$ (see text), which does not allow emission with $\vartheta_1+\vartheta_2=180^o$ for an equal energy sharing. The sum energy is plotted for $130^o\leq\vartheta_1+\vartheta_2\leq230^o$ and (Fig. \ref{fig2}(a)) equal energy sharing, (Fig. \ref{fig2}(b)) unequal energy sharing condition. Here, only one electron was measured and the momentum vector, respectively the kinetic energy, of the second electron was determined by momentum conservation and the measured ion momentum. The numbers in (a), (c), (d), and (f) correspond to the amount of absorbed photons $n$. For helium, S$_1$ and S$_2$ forbid absorption of an odd number of photons (a),(d). Due to the opposite parity of the ground state, for neon even numbers of photons are suppressed (c),(f).}
		\label{fig3}
	\end{center}
	\end {figure}
	
	As introduced above, selection rules hold true for all odd numbers of absorbed photons while for even numbers they do not apply \citep{Ni12}. As demonstrated in Fig. \ref{fig2}(a) our experimental results allow for counting how many photons have been absorbed. This prerequisite allows us to observe a possible manifestation of those selection rules for the first time in strong field double ionization. The corresponding results are shown in Fig. \ref{fig3}. In the left (right) column the occurrence of selection rule S$_1$ (S$_2$ ) is examined. Let us first consider S$_1$. In Fig. \ref{fig3}(b) the sum energy of the emitted electron pair is shown for the subset of double ionization events where the electrons are emitted approximately back-to-back  (angle $\alpha$ between both electrons is $ 140^o\leq\alpha\leq180^o$). Every single ATDI peak is observed as depicted by the black arrows in Fig. \ref{fig3}. If we now restrict this dataset to cases, where in addition both electrons share the total excess energy equally (the energy difference of both electrons is less than $1$~eV), we obtain the histogram shown in panel (a) of Fig. \ref{fig3}. For these events, selection rule (1) applies and only ATDI peaks with an even number of absorbed photons (labeled by the numbers respectively) are observed. This is in line with the prediction, that for an odd number of absorbed photons no emission within this part of phase space is allowed.\\
	
	The same approach is taken to test selection rule (2) in Fig. \ref{fig3}(d,e). Here the individual angles $\vartheta_1,\vartheta_2$ between the electrons and the polarization axis (i.e. in the laboratory frame) are now examined. As a reference, the sum energy of all electron pairs fulfilling $130^o<\vartheta_1+\vartheta_2<230^o$ is plotted in Fig. \ref{fig3}(e).  Again, all ATDI peaks are present. That changes, if one selects the subset of electrons which fulfill "equal energy sharing" in Fig. \ref{fig3}(d). As for selection rule S$_1$, also for S$_2$ only those ATDI peaks survive that are related to an even number of absorbed photons.
	
	 To illustrate the general character of the selection rules we show the equivalent data obtained employing neon as a target gas in Fig. \ref{fig3}~c,f. The ground state ($^3P$) of Ne$^{2+}$ has the opposite parity as compared to He$^{2+}$. Consequently, the absorption of odd numbers of photons is suppressed. We cannot resolve the final ionic state, but expect the contribution from the second and third excited state ($^1D$ and $^1S$) to be significantly smaller as they are 3.2 and \unit[6.9]{eV} higher in energy.
	
	In conclusion we have shown experimental evidence for two quantum mechanical selection rules shaping the two electron continuum in multiphoton double ionization. Our findings show that despite the great success of classical modeling of strong field double ionization, details of the process cannot be understood without discussing subtle consequences on the quantum nature of photon matter interaction. The consequence of the quantum nature of the process are energy quantization (see also \citep{Wu17} for a molecular case) as well as well-defined parity and angular momentum transfer. This makes double ionization a prominent example highlighting that the quantum entanglement of the electrons in atomic and molecular bound states is not broken by the interaction with a strong laser field, instead this entanglement is projected into the continuum becoming visible as electron-electron correlation in coincidence experiments.
	
	\textbf{Acknowledgment} This work was supported by DFG. K. H. and A. H. thank the Studienstiftung des deutschen Volkes for financial support. We thank A. Scrinzi, Jinzhen Zhu, Vinay Majety, U. Thumm and A. Becker for helpful discussions.
	
	\bibliographystyle{apsrev4-1}
	%\bibliography{paper}
	%

\end{document}